\documentclass[aps,pra,notitlepage,10pt,twocolumn,superscriptaddress,nofootinbib,floatfix]{revtex4-2}
\usepackage[utf8]{inputenc}
\usepackage[english]{babel}
\usepackage[T1]{fontenc}
\usepackage{amsmath}
\usepackage{amsfonts}
\usepackage{amssymb}
\usepackage{amsthm}
\usepackage{graphicx}
\usepackage{mathtools}
\usepackage[dvipsnames]{xcolor}
\usepackage{hyperref}
\hypersetup{citecolor=mpGreen}
\hypersetup{colorlinks=true}
\hypersetup{linkcolor=mpBlue}
\hypersetup{urlcolor=mpBlue}

\definecolor{mpBlue}{RGB}{21, 101, 192} 
\definecolor{mpRed}{RGB}{198, 40, 40}   
\definecolor{mpGreen}{RGB}{46, 125, 50} 
\undef\Re
\DeclareMathOperator{\Re}{Re}
\undef\Im
\DeclareMathOperator{\Im}{Im}
\DeclareMathOperator{\dd}{d \!}
\DeclareMathOperator{\e}{e}
\DeclareMathOperator{\RR}{\mathbb{R}}
\DeclarePairedDelimiter\mb{\lbrace\!\lbrace}{\rbrace\!\rbrace}

\makeatletter
\newcommand*{\bra}[1]{\langle #1 \@ifnextchar\ket{}{|}}
\makeatother
\newcommand*{\ket}[1]{| #1 \rangle}
\newcommand*{\braket}[1]{\langle #1 \rangle}
\newcommand*\ketbra[1]{\ket{#1}\!\bra{#1}}
\newcommand*{\unit}[1]{\,\mathrm{#1}}
\newtheorem*{observation*}{Observation}
\newcommand*\Paragraph[1]{\emph{#1.}---}
\begin{document}

\title{Probing the nonclassical dynamics of a quantum particle in a gravitational field}

\author{Martin Pl\'{a}vala}
\email{martin.plavala@itp.uni-hannover.de}
\affiliation{Institut für Theoretische Physik, Leibniz Universität Hannover, 30167 Hannover, Germany}

\author{Stefan Nimmrichter}
\affiliation{Naturwissenschaftlich-Technische Fakultät, Universität Siegen, 57068 Siegen, Germany}

\author{Matthias Kleinmann}
\affiliation{Naturwissenschaftlich-Technische Fakultät, Universität Siegen, 57068 Siegen, Germany}

\begin{abstract}
In quantum mechanics, the time evolution of particles is given by the Schrödinger equation. It is valid in a nonrelativistic regime where the interactions with the particle can be modelled by a potential and quantised fields are not required. This has been verified in countless experiments when the interaction is of electromagnetic origin, but also corrections due to the quantised field are readily observed. When the interaction is due to gravity, then one cannot expect to see effects of the quantised field in current-technology Earth-bound experiments. However, this does not yet guarantee that in the accessible regime, the time evolution is accurately given by the Schrödinger equation. Here we propose to measure the effects of an asymmetric mass configuration on a quantum particle in an interferometer. For this setup we show that with parameters within experimental reach, one can be sensitive to possible deviations from the Schrödinger equation, beyond the already verified lowest-order regime. Performing this experiment will hence directly test the nonclassical behaviour of a quantum particle in the gravitational field.
\end{abstract}

\maketitle

\Paragraph{Introduction}%
Quantum mechanics enables us to model physical systems and explore the behaviour of the systems with respect to those models. In the simplest cases, one needs to specify the initial state of the system, the Hamiltonian generating the time evolution, and the measurement gathering the data. With one century of experience in applying quantum mechanics, much confidence has built up in the theory, by finding great agreement between theoretical models and observations. But this story of success should not stop us to investigate whether quantum mechanics is the correct language to describe physical systems in all circumstances. This is further motivated by the fact that some of our most fundamental models of nature still have their share of open ends and inconsistencies. Although those might be overcome, it can also be worth undertaking the attempt to actively conceive experiments that can show a deviation, despite all existing evidence in favour of the theory.

Recently, low-energy table-top experiments utilizing the high precision available in quantum optical setups has come into focus for testing quantum mechanics, in particular using matter waves and nanomechanical objects. The main idea is to use the well-understood methods from quantum optics to prepare, manipulate, and measure a probe system, so that we can have high confidence that the state and the measurements at this stage are well described by quantum mechanics. While we do have a consistent and thoroughly tested quantum description of the electromagnetic interaction and their mechanical effect on matter, the gravitational interaction is an interesting candidate due to its availability in those setups. But so far, neither the gravitational attraction between macroscopic objects \cite{tan2020improvement,westphal2021measurement,fuchs2024measuring} nor the dynamics of quantum particles in a gravitational background field \cite{asenbaum2016phase,overstreet2022observation,panda2024measuring} show any hint of a deviation from Newtonian gravity in the considered mass and energy regimes. Moreover, matter-wave experiments with neutrons have also shown that linearised gravity on Earth can take part in the emergence of nonclassical behaviour \cite{nesvizhevsky2002quantum}.

\begin{figure}
\centering
\includegraphics[width=\linewidth]{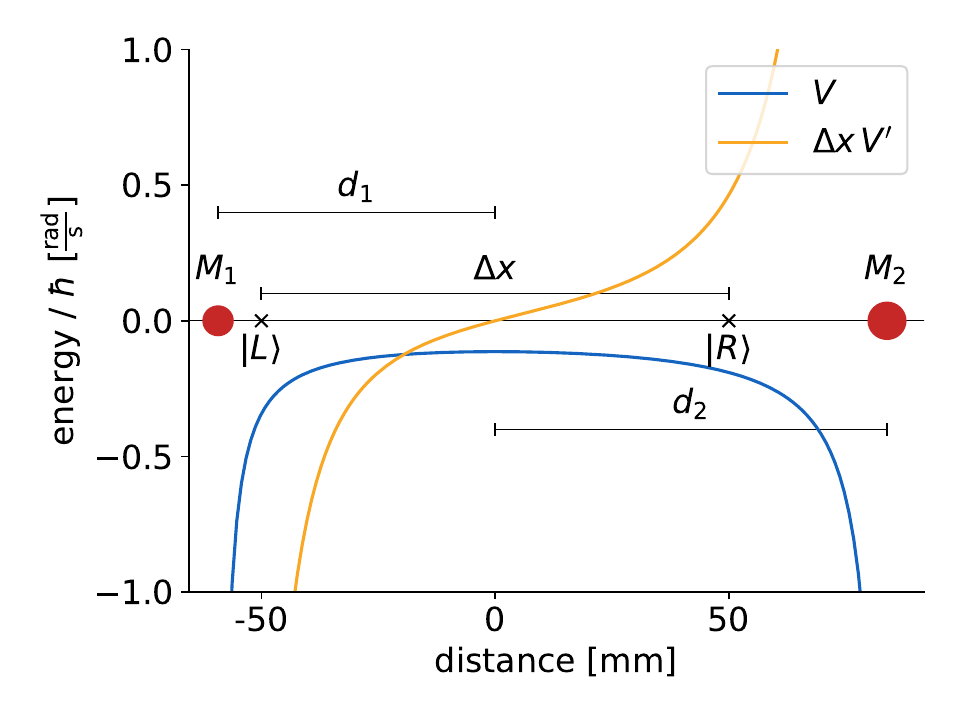}
\caption{Interferometric scheme to probe the dynamics of a quantum particle beyond the classical limit. A particle of mass $m$ is in superposition of the paths $\ket{L}$ and $\ket{R}$ with distance $\Delta x$. External masses $M_1$ and $M_2$ are at distances $d_1$ and $d_2$, respectively, from the midpoint of $\Delta x$. By choosing $M_1 / d_1^2 = M_2 / d_2^2$, the ``classical'' part of the time evolution cancels out while the reminder of the evolution leads to interference. Here a caesium atom with $m=133u$, $\Delta x = 10 \unit{cm}$ and two tungsten balls with $M_1 = 20 \unit{g}$, $M_2 = 40 \unit{g}$, $d_1 \approx 5.9 \unit{cm}$, $d_2 \approx 8.4 \unit{cm}$ are depicted in scale. Also shown are the potential energy $V(x)$ and the rescaled force $\Delta x V'(x)$.}
\label{fig:exp-scheme}
\end{figure}

How can we test a deviation of quantum mechanics if (i) the deviating theory is not yet known and (ii) the initial state and the measurements must fit within standard quantum mechanics? In this work, we seek for ``a deviation from unitarity,'' that is, a violation of the Schrödinger equation or, equivalently, the von Neumann equation, $\dot\rho=-(i/\hbar)[H,\rho]$, describing the time evolution of a density matrix $\rho$ in a system with Hamiltonian $H$. This type of time evolution has two properties that we generally demand from a fundamental dynamical law: it is time-local and linear in $\rho$. Possible alternative dynamical laws for mechanical degrees of freedom range from theories with a modified equation of motion of the form $\dot \rho=\mathcal L\rho$, such as wavefunction collapse and incoherent gravity models \cite{kafri2014classical,tilloy2016sourcing,khosla2018classical,tilloy2019does,oppenheim2023postquantum,carney2023strongly,tilloy2024general}, to general operational theories for continuous-variable systems, specified by sets of allowed preparations, time evolutions, and measurements in phase space \cite{plavala2022operational, plavala2023generalized}. We exclude here nonlinear theories such as the Schrödinger--Newton equation \cite{diosi1984gravitation, giulini2011classical}, which could allow for faster-than-light signalling \cite{gisin1990weinberg, gruca2024correlations}.

The goal of this paper is to propose an experiment measuring the nonclassical dynamics of a quantum particle in a gravitational field. To do so we consider a very generic form of the time evolution and we identify the part that represents the classical limit. This enables us to design an interferometric experiment with parameters within experimental reach, see Fig.~\ref{fig:exp-scheme}, such that the previously-identified classical term is eliminated and the yet-unknown part of the time evolution is measured. This constitutes then a test for deviations from the Schrödinger equation in quantum gravity.

\Paragraph{General time evolution in phase space}%
We choose a formulation of quantum mechanics in phase space since it is more versatile in describing nonunitary dynamics. For simplicity we restrict the discussion to one-dimensional systems, the phase space of which consists of all phase space points $(q,p)$ where $q$ represents position and $p$ represents momentum. In classical mechanics, the state of a system is described by a density function $\rho(q,p)$ that is positive, $\rho(q,p) \geq 0$, and normalized, $\int_{\RR^2} \rho(q,p) \dd q \dd p = 1$. The time evolution is given by the Liouville equation $\dot{\rho} = \{H, \rho\}$, where $H = H(q,p)$ is the Hamiltonian as a function in phase space and
\begin{equation}
 \{f,g\} = f (\overleftarrow{\partial_q} \overrightarrow{\partial_p} - \overleftarrow{\partial_p} \overrightarrow{\partial_q}) g = \frac{\partial f}{\partial q} \frac{\partial g}{\partial p} - \frac{\partial f}{\partial p} \frac{\partial g}{\partial q}
\end{equation}
is the Poisson bracket. 

The state of a quantum system in phase space is described by the Wigner quasiprobability distribution
\begin{equation}
 W(q, p) = \dfrac{1}{2 \pi \hbar} \int_{\RR} \e^{-i p s/\hbar} \bra{x-\frac{s}{2}} \hat\rho \ket{x+\frac{s}{2}} \dd s,
\end{equation}
where $\langle x|\hat\rho|y\rangle$ is the density matrix of the system state in position representation. While $W$ is still normalized, $\int_{\RR^2} W(q,p) \dd q \dd p = 1$, it is in general not positive anymore. The time evolution of the Wigner function is given by the time evolution of the density matrix according to the von Neumann equation. This induces a time evolution of the Wigner function \cite{groenewold1946principles, moyal1949quantum} which satisfies $\dot W=\mb{H,W}$ where
\begin{multline}
\mb{H, W} = \{H, W\} \\
+ \sum_{n=1}^\infty \dfrac{(-1)^n \hbar^{2n}}{2^{2n} (2n+1)!} H \left( \overleftarrow{\partial_q} \overrightarrow{\partial_p} - \overleftarrow{\partial_p} \overrightarrow{\partial_q} \right)^{2n+1} W
\label{eq:W_Moyal}
\end{multline}
is the Moyal bracket. In the classical limit, the Wigner function should become a positive distribution and the Moyal bracket should reduce to the Poisson bracket; it is suggestive to do so by using the limit $\hbar \to 0$, although this is not always straightforward \cite{case2008wigner}.

Our aim is to single out the form of the time evolution under gravity in the regime of table-top experiments \cite{aspelmeyer2022zeh}. In such a setting, initial state preparations and final measurements are performed using electromagnetic interactions. We therefore require that the initial and the final system states are consistent with quantum mechanics and represented by valid density matrices. Gravity affects the time evolution in between, which is thus the only aspect of a theory of gravity that is accessible in this scenario.

Rather than considering a particular theory of gravity, we consider a large class of theories and devise experimental schemes that narrow down the theories within this class. The class that we consider will be as general as possible and includes arbitrary linear time evolution. We characterize the class by time evolutions of the general form \cite{plavala2023generalized}
\begin{equation} \label{eq:timeEvo-phaseSpace}
\dot{W} = \{H, W\} + G_H(W),
\end{equation}
where $G_H$ can be any differential operator (with arbitrarily many terms) acting on $W$ and depending on $H$. We also naturally assume that the resulting time evolution is well defined and produces only positive probabilities for the measurement outcomes. This, together with other assumptions can be used to put some restrictions on the form of $G_H$ \cite{jiang2024unification, jiang2024framework}. Of course, the class includes the standard theory in which gravity acts as a Newtonian potential in the Schrödinger equation. Moreover, we expect that the $G_H$ becomes negligible in the classical limit, so that Eq.~\eqref{eq:timeEvo-phaseSpace} reduces to the Liouville equation and differences between theories are no longer observable. We thus formalize our development so far as follows:
\begin{observation*}
In the regime of table-top experiments, initial and final states of the system must be described by density matrices. Gravity affects only the intermittent time evolution. Assuming the time evolution is linear, it is in general of the form given by \eqref{eq:timeEvo-phaseSpace}.
\end{observation*}

Our decomposition of the time evolution is reminiscent of the gravitational Aharonov-Bohm effect, proposed in \cite{hohensee2012force} and observed in \cite{overstreet2022observation}. The crucial difference is that modifications of quantum theory based on time evolutions other than the Schr\"{o}dinger equation could predict the gravitational Aharonov-Bohm effect. The specific form of the generator of time evolution $G_H$ fully specifies the effective theory in the considered regime, enabling us to predict the outcomes of other future experiment in the same regime. Our aim will be to design experiments to directly measure the operator $G_H$ and thus fully characterize the quantum features of gravity in the table-top regime.

The main hurdle in observing the contributions of $G_H$ to the total time evolution is that in many experimental setups, the Poisson bracket $\{H, W\}$ acts as the leading term while $G_H(W)$ contributes only minor corrections. For this reason, we propose an interferometer configuration with two source masses in which the contribution of the Poisson bracket to the interference signal vanishes, allowing us to directly observe the effect of the nonclassical term. In particular, we get an observable effect when gravity acts on quantum matter via the Schrödinger equation, that is, when $G_H(W)$ is given by the second line in Eq.~\eqref{eq:W_Moyal}.

\Paragraph{Interferometer scheme}%
We consider the simplest Mach--Zehnder interferometer scheme in which a quantum particle (say, an atom or a molecule) is split uniformly into two arms and kept in a superposition of two statically distinct states $\ket{L}$ and $\ket{R}$ separated by the distance $\Delta x$. The states shall be given by well-localized Gaussian wave packets that are approximately orthogonal, $\braket{L|R} \approx 0$, and they shall be held at fixed positions $x = \pm \Delta x/2$ in a static gravitational potential $V(x)$ sourced by two nearby macroscopic masses for the duration $t$ of the superposition, see Fig.~\ref{fig:exp-scheme}. In this effectively one-dimensional description, $x$ is the coordinate along which the two wave packets are placed. 

For experimental feasibility, we refer to the lattice interferometer setup of Ref.~\cite{xu2019probing,panda2024coherence} in which an ensemble of caesium atoms were kept in a superposition over several micrometers for up to 70 seconds. Alternatively, the experiment could be performed in free fall as observable phase shift would be accumulated already after 1 second.
We shall neglect the short time it takes to split and recombine the two arms for now and describe the initial particle state by $\hat{\rho}(0) = \ketbra{+}$. A more realistic description in which the the masses are also present during the splitting and recombination stage will be considered later. The measurement signal $S$ is given by a projection of the final particle state $\hat{\rho} (t)$ onto the superposition $\ket{\varphi} = (\ket{L} + e^{-i\varphi}\ket{R})/\sqrt{2}$, with a suitably chosen recombination phase $\varphi$ for maximum sensitivity to gravitational phase shifts,
\begin{equation} \label{eq:interferomenter-St}
    S(t) \propto \braket{+|\hat\rho (t)|+} = \frac{1}{2} + \Re\left[ \braket{L|\hat\rho(t) e^{-i\varphi}| R} \right] .
\end{equation}

We make two additional simplifying assumptions: (i) the free dispersion of each wave packet can be omitted for the duration of the experiment, and (ii) the variation of the potential $V(x)$ and thus its kinematic effect over the spread of each individual wave packet is negligible. The first assumption is satisfied if the wave packet is kept trapped in the well of an optical lattice as in Ref.~\cite{xu2019probing}, but it could be extended to the case of free-falling wave packets by defining time-dependent states $\ket{L(t)}$ and $\ket{R(t)}$ that incorporate the free evolution in Earth's linear gravity. The second assumption is uncritical and often made in interferometric gravimetry; in the quantum case, it amounts to approximating the influence of $V(x)$ by a relative phase between the two arms that accumulates over time. This and the corresponding approximations for more general $G_H$ are detailed below. Both assumptions together imply that the final state $\hat\rho(t)$ remains in the two-dimensional Hilbert space spanned by $\ket{L},\ket{R}$, and the only non-negligible dynamics is that of the four respective matrix elements.

We have now two options to compute the predictions of the experiment: we can follow the approach recently used in Ref.~\cite{marchese2024newton} and map the initial state $\hat\rho(0)$ to the respective Wigner function and apply Eq.~\eqref{eq:timeEvo-phaseSpace} to it, or we can use the Weyl quantisation to express Eq.~\eqref{eq:timeEvo-phaseSpace} as an equation for the density matrix. We will opt for the latter method as it gives clear and easily computable predictions. Given a Wigner quasiprobability distribution $W(q,p)$, one recovers the respective density matrix $\hat{\rho}$ in the position representation as \cite{hall2013quantum}
\begin{equation}
\bra{x} \hat{\rho} \ket{y} = \int_{\RR} \e^{i p (x-y)/\hbar} W \left(\frac{x+y}{2}, p\right) \dd p.
\end{equation}
The time evolution as stated in Eq.~\eqref{eq:timeEvo-phaseSpace} has two terms on the right hand side: the first corresponds to the classical contribution of the Poisson bracket, while the other term is the---in principle unknown---operator $G_H$. We denote the Weyl quantisation of $G_H(W)$ as $\mathcal{G}_H(\hat{\rho},x,y)$ where $\mathcal{G}_H$ is a suitable superoperator. Assuming the standard form of the Hamiltonian $H(q,p) = \frac{p^2}{2m} + V(q)$, we get
\begin{multline}
\braket{x|\dot{\hat{\rho}}|y} =  \frac{1}{i \hbar} \braket{x| \left[\frac{\hat{p}^2}{2m}, \hat{\rho} \right] |y} + \frac{x - y}{i \hbar} V'\left(\frac{x + y}{2}\right) \braket{x|\hat{\rho}|y}  \\
+ \mathcal{G}_H(\hat{\rho},x,y),
\label{eq:timeEvo-operator}
\end{multline}
where $V' = \dd V/\dd x$ is the derivative of the potential; see the Appendix~\ref{appendix:derivationTimeEvo} for the full derivation of Eq.~\eqref{eq:timeEvo-operator}. Note that, depending on $\mathcal{G}_H$, the time evolution can be nonunitary, and we merely require that it results in valid states $\hat\rho(t)$. We again formalize our progress so far:
\begin{observation*}
The effective theory of how gravity influences single delocalized quantum system is fully described by the unknown term $\mathcal{G}_H(\hat{\rho},x,y)$ that corresponds to the non-classical effects of the time evolution. If this term is known, then one can predict the outcome of any other experiment in the same regime.
\end{observation*}
One can use methods developed in quantum information for estimating the generators of time evolution \cite{boulant2003robust,howard2006quantum,mazza2021machine,samach2022lindblad,gebhart2023learning,hangleiter2024robustly,brand2024markovian,wallace2024learning} to directly measure the unknown term $\mathcal{G}_H(\hat{\rho},x,y)$. This would involve performing several experiments, with initial state prepared not only in the $\ket{+}$, but also the localized states $\ket{L}$ and $\ket{R}$, and the state $\frac{1}{\sqrt{2}}(\ket{L} + i \ket{R})$. Also not only the overlap with the $\ket{\varphi}$ state should be measured, but with all the aforementioned states in different runs of the experiment in order to perform tomography of the time evolution and thus experimentally verify the form of the unknown generator $\mathcal{G}_H(\hat{\rho},x,y)$. That being said, it is reasonable to expect that the most significant experiment would be the one in which the initial state is $\ket{+}$ and we would observe a phase shift predicted by the Schr\"{o}dinger equation or by other previously investigated models. Hence we concentrate on this scenario.

With the help of our simplifying assumptions (i) and (ii), we obtain a differential equation for the effectively two-dimensional state. According to (i), we can neglect the kinetic energy term in Eq.~\eqref{eq:timeEvo-operator}. Assumption (ii) means that we may replace the coordinates $x$ and $y$ in the $V'$-term by their respective mean values over each wave packet. Let us assume further that (iii) the same holds for the yet unspecified term $\mathcal{G}_H$: it does not appreciably affect the trajectory of a well-localised atomic-scale wave packet. For the diagonal matrix elements $\bra{L} \hat\rho\ket{L} = \int_{\RR} \bra{x} \hat\rho\ket{y} \bra{L}\ket{x} \bra{y}\ket{L} \dd x \dd y$ and $\bra{R} \hat\rho\ket{R}$, we can set $x\approx y \approx \pm \Delta x/2$, which implies that the $V'$-term does not contribute. For the coherence term $\bra{L} \hat\rho\ket{R}$, we have $x \approx -\Delta x/2$ and $y \approx \Delta x/2$. Putting everything together, we arrive at
\begin{align}\label{eq:eom_2state}
    \bra{L} \dot{\hat{\rho}} \ket{L} &\approx \mathcal{G}_H(\hat{\rho},-\frac{\Delta x}{2},-\frac{\Delta x}{2}) \approx - \bra{R} \dot{\hat{\rho}} \ket{R}, \\
    \bra{L} \dot{\hat{\rho}} \ket{R} &= \overline{\bra{R} \dot{\hat{\rho}} \ket{L}} \approx i \omega_C \bra{L} \hat{\rho} \ket{R} + \mathcal{G}_H(\hat{\rho},-\frac{\Delta x}{2},\frac{\Delta x}{2}) .\nonumber
\end{align}
Here we denote by $\omega_C = \Delta x V'(0)/\hbar$ the only remaining contribution of the classical Poisson bracket to the time evolution.

Should the time evolution be governed by the standard Schrödinger equation, then under assumption (i), we simply have $\bra{x} \dot{\hat{\rho}} \ket{y} \approx [V(x)-V(y)]\bra{x}\hat{\rho}\ket{y}/i\hbar$. By virtue of (ii), this results in a pure phase between the two arms,
\begin{equation}\label{eq:eom_2state_Q}
  \text{Schrödinger:}\quad
  \begin{array}{l}
  \bra{L} \dot{\hat{\rho}} \ket{L} \approx \bra{R} \dot{\hat{\rho}} \ket{R} \approx 0 \\
  \bra{L} \dot{\hat{\rho}} \ket{R} \approx i\omega_Q \bra{L} \hat{\rho} \ket{R}
  \end{array}
\end{equation}
with $\hbar \omega_Q = V(\Delta x/2)-V(-\Delta x/2)$. The corresponding interference signal \eqref{eq:interferomenter-St} is a perfect fringe oscillation at frequency $\omega_Q$,
\begin{equation} \label{eq:interferomenter-St_Q}
   \text{Schrödinger:}\quad S(t) \propto \frac{1+\cos (\omega_Q t-\varphi)}{2}.
\end{equation}

Hence, in the two-arm approximation regime of assumptions (i) and (ii), the net effects of the classical Poisson bracket and of the quantum Moyal bracket are phase oscillations at the respective frequencies $\omega_C$ and $\omega_Q$.

\Paragraph{Source mass configuration}%
We now consider the setting depicted in Fig.~\ref{fig:exp-scheme} in which two macroscopic spherical source masses $M_1$ and $M_2$ are situated in the vicinity of the interfering particle of mass $m$, the centres of mass all lined up on the $x$-axis to facilitate a one-dimensional treatment. The mass $M_1$ is placed at $x=-d_1$ left of the particle and $M_2$ at $x=d_2$ right of the particle. In between at $|x| < d_1,d_2$, the gravitational potential on the particle and its derivative read as
\begin{align}
   V(x) &= -\dfrac{GmM_1}{d_1 + x} - \dfrac{GmM_2}{d_2 - x}, \\
   V'(x) &= \dfrac{GmM_1}{(d_1 + x)^2} - \dfrac{GmM_2}{(d_2 - x)^2},
\end{align}
with $G$ the gravitational constant; here we assume the standard gravitational Newtonian potential, but this can also be experimentally tested at the relevant scale \cite{cheng2024proposal}. The corresponding Poisson and Moyal phase frequencies in the approximate two-state evolution of the particle are
\begin{align}
    \omega_C &= \dfrac{Gm \Delta x}{\hbar} \left( \frac{M_1}{d_1^2} - \frac{M_2}{d_2^2} \right), \\
    \omega_Q &= \dfrac{Gm \Delta x}{\hbar} \left( \frac{M_1}{d_1^2 - \frac{\Delta x^2}{4}} - \frac{M_2}{d_2^2 - \frac{\Delta x^2}{4}} \right).
\end{align}
With two masses instead of a single one, we can now choose a configuration that clearly identifies the phase shift with angular frequency $\omega_C$ caused by the Poisson bracket, distinguishing it from the quantum Moyal case at $\omega_Q$ or any other nonclassical phase shift caused by the unknown term $\mathcal{G}_H$ in Eq.~\eqref{eq:eom_2state}. In particular, by placing two different masses ($M_1\neq M_2$) asymmetrically such that $M_1 / d_1^2 = M_2 / d_2^2$, we achieve $\omega_C = 0 \neq \omega_Q$ for any arm separation $\Delta x>0$ and thus sensitivity to any nonclassical deviation. Alternatively, we could seek deviations from the quantum case by adjusting for $\omega_Q = 0$ or vary the mass positions in subsequent repetitions to probe the dependence of the phase shift on the distances $d_1, d_2$. An experimentally feasible configuration for $\omega_C=0$, but with a significant quantum phase shift, is given below.

At this point, we compare our setup with previous experiments. First, the phase shift due to the gravitational field of Earth observed in Ref.~\cite{colella1975observation} provides only limited information about the unknown term $\mathcal{G}_H(\hat{\rho},x,y)$. To see this explicitly, let $M_1$ represent Earth and $M_2=0$. Any realistic arm separation will always be orders of magnitude smaller than Earth's radius, $\Delta x \ll d_1$, which results in a linear potential and $\omega_C \approx \omega_Q$, so the phase shift can be attributed to the Poisson bracket.

Second, a recent interference experiment with Rubidium atoms \cite{overstreet2022observation} has probed the gravitational Aharonov--Bohm effect of a macroscopic source mass that is brought in close vicinity to one of the interferometer arms. Concretely, the measurement data has distinguished the interferometric phase shift caused by the full Newtonian potential against one in which gravity enacts a different Newtonian acceleration along each arm. However, although the experimental setup has similarities with the setup proposed below, it is tailored to refuting the hypothesis of local Newtonian acceleration, which is different from the phase shift generated by the Poisson bracket. This calls for dedicated measurements and data analysis to probe the nature of the gravitational phase shift in this setup and seek for possible deviations from $\omega_Q$ and in general from the Schr\"{o}dinger equation, as predicted by, for example, measurement-feedback models \cite{kafri2014classical,tilloy2016sourcing,khosla2018classical,tilloy2019does,oppenheim2023postquantum,carney2023strongly,tilloy2024general,trillo2024diosi} or other semiclassical models \cite{hall2018two,doner2022gravitational,gonzalez2023mixed,gollapudi2024state} of gravity.

\Paragraph{Experimental parameters}%
We now propose a feasible experimental setting that would be able to measure the effect of the unknown superoperator $\mathcal{G}_H$ on the interferometric phase. The natural choice of time evolution, the Moyal bracket, shall yield a measurable phase shift $\omega_Q t$, whereas the Poisson bracket contribution shall vanish, $\omega_C = 0$.

\begin{figure}
\centering
\includegraphics[width=\linewidth]{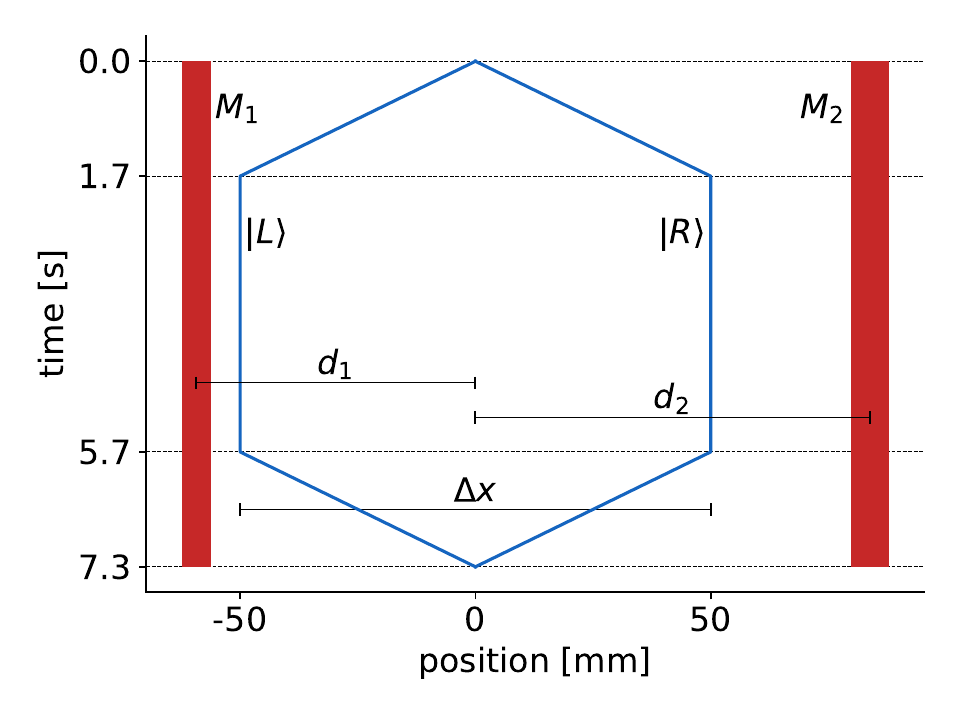}
\caption{Spacetime diagram of the interferometric setup. At time $t=0$, a coherent momentum splitting by $\pm 10\hbar k$ is applied to the atom, yielding two arms at a relative velocity of $0.6 \unit{cm / s}$. After $1.67 \unit{s}$, another pulse eliminates the relative motion, and the arms are kept at a constant separation of $10\unit{cm}$ for $4 \unit{s}$. Then the pulse sequence is applied in reverse order to recombine the arms. For the proposed mass configuration, the phase shift would vanish exactly if the systematic effect of gravity were described by the Poisson bracket. The Schrödinger equation predicts a phase shift of $0.675 \unit{rad}$.}
\label{fig:exp-spacetime}
\end{figure}

We consider a single caesium atom ($m=133 \unit{u}$) in spatial superposition with $\Delta x = 10 \unit{cm}$ and two tungsten balls of masses $M_1 = 20 \unit{g}$ and $M_2 = 40 \unit{g}$ as the nearby sources of gravity, see Fig.~\ref{fig:exp-scheme}. Greater arm separations were already demonstrated in a free-fall experiment \cite{kovachy2015quantum}. To be concrete, we envisage the following interferometer protocol: an initial control pulse splits the atom state coherently into two arms separated by $\pm 10\hbar k$ in momentum. Assuming a laser wavelength $\lambda \approx 1 \unit{\mu m}$, the arms diverge at a relative velocity of $6\,\unit{cm / s}$. A second pulse eliminates the relative motion after $1.7 \unit{s}$, when the desired arm separation $\Delta x$ is reached. The atom is kept in this superposition for $4\unit{s}$, subject to the gravitational potential of the two masses. For recombination, the initial pulse sequence is performed in reverse order, after which the atom is detected in either output state. A spacetime diagram of the experiment is depicted in Fig.~\ref{fig:exp-spacetime}.

By placing the left tungsten ball $M_1$ such that its surface is $3\unit{mm}$ away from the left interferometer arm, we can achieve a significant gravitational effect and safely ignore other systematic effects such as the influence of the Casimir--Polder interaction or of surface roughness \cite{panda2024measuring}. The right ball is placed further away from the right arm such that $\omega_C=0$ during the whole duration of the experiment. This is possible because, even though the distance of the interferometer arms depends on time, the center line in between the arms does not move during the experiment. It follows that the distances $d_1$ and $d_2$ of the masses from that center line do not change as well. We can thus place them such that $M_1 / d_1^2 = M_2 / d_2^2$ and thus $\omega_C = 0$ holds at all times, eliminating the systematic effect that would be present in experiments with just one external mass. Given the two ball radii, $R_1 \approx 6.3 \unit{mm}$ and $R_2 \approx 7.9 \unit{mm}$, the distances are $d_1 = \frac{\Delta x}{2} + R_1 + 1 \unit{mm} \approx 5.9 \unit{cm}$ and $d_2 = d_1 \sqrt{M_2 / M_1} \approx 8.4 \unit{cm}$. For the phase shift induced by the Moyal bracket, we then get
\begin{equation}
\omega_Q \approx 0.152 \frac{\unit{rad}}{\unit{s}},
\end{equation}
or $0.61 \unit{rad}$ after $4 \unit{s}$, which one should be able to resolve in state-of-the-art setups \cite{overstreet2022observation,panda2024measuring}. Small phase shifts are best discerned at a recombination phase $\varphi = \pm \pi/2$.

The gravitational phase shift also accumulates during the splitting and the recombination stage, when the arm separation $\Delta x$ and thus also the angular frequency $\omega_Q$ are time dependent. Solving the differential equation in \eqref{eq:eom_2state_Q} then results in
\begin{equation}\label{eq:eom_2state_timeDep}
\bra{L} {\hat{\rho}}(t) \ket{R} \approx \e^{i \int_0^t \omega_Q(\tau) \dd \tau} \bra{L} \hat{\rho}(0) \ket{R},
\end{equation}
which can be integrated analytically for the scheme depicted in Fig.~\ref{fig:exp-spacetime}, or otherwise numerically. For our proposed parameter settings, the splitting and recombination stages increase the gravitational phase shift by less than $0.07 \unit{rad}$.

We remark that, in order to calculate the actual gravitational phase shifts and find a configuration of source masses that leads to $\omega_C \approx 0$ in a realistic experiment, one should take into account the exact geometry of the masses and the particle trajectory in three dimensions. For the sake of conceptual clarity, we have omitted these details and postpone such an in-depth analysis to future experimental case studies.

\Paragraph{Probing gravity beyond the Moyal bracket}%
As of now, a general and consistent quantum field theory of gravity is lacking. Quite possibly, such a theory could lead to a unitary description of the gravitational potential at low energies, guided by the Schrödinger equation. In contrast, recently proposed alternative models treat gravity as a classical field that interacts with quantum matter via an effective measurement-feedback process \cite{kafri2014classical, tilloy2016sourcing, khosla2018classical, tilloy2019does, oppenheim2023postquantum, carney2023strongly, feng2023conservation, tilloy2024general}, which results in a nonunitary dynamics. Such models would leave it open whether (nonrelativistic) gravity acts in the form of a noisy potential or a noisy classical interaction.

Our proposed interferometer setup allows one to test the validity of the Schrödinger equation in nonrelativistic gravity as well as more general hypotheses resulting in a phase shift determined neither by $\omega_C$ nor $\omega_Q$ and is sensitive to effects such as decoherence. For such a broader sensitivity, the  experiment would have to be repeated with a suitable placement of the two source masses, different initial states, final measurements, and interaction times in order to estimate the general superoperator $\mathcal{G}_H$ \cite{boulant2003robust,howard2006quantum,mazza2021machine,samach2022lindblad,gebhart2023learning,hangleiter2024robustly,brand2024markovian,wallace2024learning}. Given such an estimate, one can assess whether existing theories of gravity agree with it or otherwise infer a suitable physical model.

We now investigate a simple modification of the superoperator $\mathcal{G}_H$, a more general discussion is deferred to the Appendix~\ref{appendix:generalGH}. Specifically, we use the Tilloy--Diósi model of measurement-feedback gravity \cite{tilloy2016sourcing} as an instructive example. It is based on a model of self-gravitational collapse \cite{diosi1989models,penrose1996gravity}, reinterpreted as a continuous mass density measurement and augmented by a Newtonian feedback potential. Crucially, the model requires a regularisation of the mass density to avoid divergences, and in our case this leads to a model-dependent modified phase shift $\omega_G \neq \omega_C,\omega_Q$. This is accompanied by a diffusion process that mainly causes the two interferometer arms to dephase at a rate $\lambda$, leading to the time evolution
\begin{equation}\label{eq:timeEvo-TilloyDiosyPenrose}
  \text{Tilloy--Diósi:}\quad
  \begin{array}{l}
  \bra{L} \dot{\hat{\rho}} \ket{L} \approx \bra{R} \dot{\hat{\rho}} \ket{R} \approx 0 \\
  \bra{L} \dot{\hat{\rho}} \ket{R} \approx (- \lambda + i \omega_G) \bra{L} \hat{\rho} \ket{R}
  \end{array}
\end{equation}
It was recently argued that, depending on the regularisation, the dephasing could be sufficiently suppressed so as to still mediate entanglement between two simultaneously interfered massive particles \cite{trillo2024diosi}. Solving Eq.~\eqref{eq:timeEvo-TilloyDiosyPenrose} for our setting results in the interference signal
\begin{equation} \label{eq:signal-TilloyDiosy}
    \text{Tilloy--Diósi:}\quad S(t) = \frac{1 + \e^{- \lambda t} \cos(\omega_G t-\varphi)}{2},
\end{equation}
where the fringe contrast decays exponentially at the dephasing rate. In practice, it would be difficult to verify the existence of this dephasing since it competes with conventional noise sources and environmental decoherence, which also contribute to the loss of fringe contrast. Regardless, the model-specific phase shift could be estimated by measuring the fringe period.

Let us finally consider a more general situation where $\bra{L} \dot{\hat{\rho}} \ket{L} = \mu_1 (\bra{L} \hat{\rho} \ket{R} + \bra{R} \hat{\rho} \ket{L}) + i \mu_2 (\bra{L} \hat{\rho} \ket{R} - \bra{R} \hat{\rho} \ket{L})$ with real numbers $\mu_1, \mu_2$. We get the same measurement signal as in Eq.~\eqref{eq:signal-TilloyDiosy} independent of $\mu_1$, $\mu_2$. We can nevertheless solve the respective differential equations, the result is a shift of the population difference between the arms proportional to $\mu_1$ and $\mu_2$; if for simplicity take $\mu_1 = \mu_2 = \mu$, then we get $\lim_{t\to\infty}\bra{L} \hat{\rho} \ket{L} = \frac{1}{2} - \frac{\mu (\omega_G - \lambda)}{\lambda^2 + \omega_G^2}$. Probing hypothetical model values for $\mu$ (or $\mu_1 \neq \mu_2$) would then require monitoring changes in the single-arm occupation.

In summary, our examples demonstrate that state-of-the-art atom interferometers are suitable platforms for probing quantum signatures of gravity in the nonrelativistic domain. By leveraging their large arm separations and long interference times, we can devise setups with external source masses that are highly sensitive to possible deviations from the canonical treatment of gravity as a $1/r$-potential in the Schrödinger equation.

\Paragraph{Acknowledgments}%
We are thankful to Igor Pikovski and Philipp Haslinger for discussions.
MP acknowledges support from the Niedersächsisches Ministerium für Wissenschaft und Kultur. This work was supported by
the Deutsche Forschungsgemeinschaft (DFG, German Research Foundation, project numbers 447948357 and 440958198
the Sino-German Center for Research Promotion (Project M-0294), and the German Ministry of Education and Research (Project QuKuK, BMBF Grant No.\ 16KIS1618K).

\bibliography{citations}

\onecolumngrid
\appendix

\section{Derivation of Equation~\eqref{eq:timeEvo-operator}} \label{appendix:derivationTimeEvo}
We need to compute the Weyl quantisation in the position representation of $\{H, W\}$ for $H(q,p) = \frac{p^2}{2m} + V(q)$. Since the Poisson bracket and Weyl quantisation are both linear, we can compute the result for each term separately. The Weyl quantisation of $\{\frac{p^2}{2m}, W\}$ is rather straightforward as for quadratic functions the Poisson and Moyal brackets coincide, that is $\{\frac{p^2}{2m}, W\} = \mb{\frac{p^2}{2m}, W}$, and since the Moyal bracket is the equivalent of commutator for Wigner functions \cite{groenewold1946principles,moyal1949quantum}, we get that the Weyl quantisation of $\{\frac{p^2}{2m}, W\}$ is $[\frac{\hat{p}^2}{2m}, \hat{\rho}]$.

The Weyl quantisation of $\{V, W\}$ yields the second term in Eq.~\eqref{eq:timeEvo-operator}. We have
\begin{equation}
\{V, W\}(q,p) = V'(q) \partial_p W(q,p)
\end{equation}
where we are using the shorthand $V'$ to denote the derivative of $V$ and $\partial_p W = \frac{\partial W}{\partial p}$. We then have
\begin{equation}
\int_{\RR} \e^{\frac{i p (x-y)}{\hbar}} \{V, W\}(\frac{x+y}{2}, p) \dd p = V'(\frac{x+y}{2}) \int_{\RR} \e^{\frac{i p (x-y)}{\hbar}} \partial_p W(\frac{x+y}{2}, p) \dd p.
\end{equation}
The Weyl quantisation of $W$ gives back $\rho$, and hence partial integration gives the desired result. Explicitly, by using the definition of the Wigner function, $W(q, p) = \dfrac{1}{2 \pi \hbar} \int_{\RR} \e^{-\frac{i p s}{\hbar}} \bra{q + \frac{s}{2}} \hat{\rho} \ket{q - \frac{s}{2}} \dd s$, we get
\begin{align}
\int_{\RR} \e^{\frac{i p (x-y)}{\hbar}} \{V, W\}(\frac{x+y}{2}, p) \dd p &= \dfrac{V'(\frac{x+y}{2})}{2 \pi \hbar} \int_{\RR^2} \e^{\frac{i p (x-y)}{\hbar}} \dfrac{\partial}{\partial p} \left( \e^{-\frac{i p s}{\hbar}} \bra{\frac{x+y+s}{2}} \hat{\rho} \ket{\frac{x+y-s}{2}} \right) \dd p \dd s \\
&= \dfrac{V'(\frac{x+y}{2})}{2 i \pi \hbar^2} \int_{\RR^2} s \e^{\frac{i p (x-y-s)}{\hbar}} \bra{\frac{x+y+s}{2}} \hat{\rho} \ket{\frac{x+y-s}{2}} \dd p \dd s \\
&= \dfrac{V'(\frac{x+y}{2})}{i \hbar} \int_{\RR} s \delta(x-y-s) \bra{\frac{x+y+s}{2}} \hat{\rho} \ket{\frac{x+y-s}{2}} \dd s \\
&= \dfrac{x - y}{i \hbar} V'(\frac{x+y}{2}) \bra{x} \hat{\rho} \ket{y}
\end{align}
which is the sought expression.

\section{Modified time evolution} \label{appendix:generalGH}
In the most general case $\mathcal{G}_H$ is an unknown superoperator and so we cannot easily obtain any predictions for the outcome of the experiment, but we will show that this is possible if $\mathcal{G}_H$ is assumed to be linear and consistent with classical mechanics. While linearity is an assumption that can be either justified by noting that nonlinear modifications of quantum mechanics are often problematic \cite{gisin1990weinberg}, one can also see linear the case of $\mathcal{G}_H$ as a first-order approximation of a nonlinear time evolution. Consistency with classical mechanics will mean that the path-localized states $\ket{L}$ and $\ket{R}$ will follow the respective classical paths, moreover this assumption is experimentally testable.

From the normalization of the density matrix $\bra{L} \hat{\rho} \ket{L} + \bra{R} \hat{\rho} \ket{R} = 1$ we get
\begin{equation}
\mathcal{G}_H(\hat{\rho},-\frac{\Delta x}{2},-\frac{\Delta x}{2}) = - \mathcal{G}_H(\hat{\rho},\frac{\Delta x}{2},\frac{\Delta x}{2})
\end{equation}
and from the hermiticity of the density matrix we get
\begin{equation}
    \mathcal{G}_H(\hat{\rho},-\frac{\Delta x}{2},-\frac{\Delta x}{2}) \in \RR
\end{equation}
and
\begin{equation}
    \overline{\mathcal{G}_H(\hat{\rho},\frac{\Delta x}{2},-\frac{\Delta x}{2})} = \mathcal{G}_H(\hat{\rho},-\frac{\Delta x}{2},\frac{\Delta x}{2})
\end{equation}
where $\overline{z}$ is the complex conjugate of $z$, so it follows that we only need to investigate $\mathcal{G}_H(\hat{\rho},\frac{\Delta x}{2},-\frac{\Delta x}{2})$ and $\mathcal{G}_H(\hat{\rho},\frac{\Delta x}{2},\frac{\Delta x}{2})$. Assuming that $\mathcal{G}_H$ is linear we get
\begin{align}
    \mathcal{G}_H(\hat{\rho},-\frac{\Delta x}{2},-\frac{\Delta x}{2}) &= a_{LL} \bra{L} \hat{\rho} \ket{L} + a_{LR} \bra{L} \hat{\rho} \ket{R} + a_{RL} \bra{R} \hat{\rho} \ket{L} + a_{RR} \bra{R} \hat{\rho} \ket{R} \label{eq:GH-a} \\
    \mathcal{G}_H(\hat{\rho},\frac{\Delta x}{2},-\frac{\Delta x}{2}) &= b_{LL} \bra{L} \hat{\rho} \ket{L} + b_{LR} \bra{L} \hat{\rho} \ket{R} + b_{RL} \bra{R} \hat{\rho} \ket{L} + b_{RR} \bra{R} \hat{\rho} \ket{R} \label{eq:GH-b}
\end{align}
where all coefficients are complex numbers.

Consistency with classical mechanics in an experimentally testable assumption that requires that $\ket{L}$ and $\ket{R}$ will follow the respective classical paths, meaning that we must have $\mathcal{G}_H(\ketbra{L},x,x) = \mathcal{G}_H(\ketbra{R},x,x) = 0$ since this is equivalent to $\bra{L} \dot{\hat{\rho}} \ket{L} = \bra{R} \dot{\hat{\rho}} \ket{R} = 0$ for $\hat{\rho} \in \{ \ketbra{L}, \ketbra{R} \}$. But then positive semidefinitness of the density matrix implies that the off-diagonal elements must be also zero at all times, which yields $\mathcal{G}_H(\ketbra{L},x,y) = \mathcal{G}_H(\ketbra{R},x,y) = 0$. Plugging these expression into \eqref{eq:GH-a} and \eqref{eq:GH-b} yields $a_{LL} = a_{RR} = b_{LL} = b_{RR} = 0$, reducing the number of the coefficient to half.

We thus get the differential equations
\begin{align}
    \bra{L} \dot{\hat{\rho}} \ket{L} = a_{LR} \bra{L} \hat{\rho} \ket{R} + a_{RL} \bra{R} \hat{\rho} \ket{L}, \label{eq:GH-LL} \\
    \bra{L} \dot{\hat{\rho}} \ket{R} = b_{LR} \bra{L} \hat{\rho} \ket{R} + b_{RL} \bra{R} \hat{\rho} \ket{L}. \label{eq:GH-LR}
\end{align}
Requiring that $\mathcal{G}_H(\hat{\rho},-\frac{\Delta x}{2},-\frac{\Delta x}{2}) \in \RR$ leads to $\Re(a_{LR}) = \Re(a_{RL})$ and $\Im(a_{LR}) = - \Im(a_{RL})$, where $\Re(z)$ is the real part of $z$ and $\Im(z)$ is the imaginary part of $z$. The equation \eqref{eq:GH-LL} then becomes
\begin{equation} \label{eq:GH-LL-final}
    \bra{L} \dot{\hat{\rho}} \ket{L} = 2 \left( \Re(a_{LR}) \Re(\bra{L} \hat{\rho} \ket{R}) - \Im(a_{LR}) \Im(\bra{L} \hat{\rho} \ket{R}) \right)
\end{equation}
and in the main text we used the notation $\Re(a_{LR}) = \mu_1$ and $\Im(a_{LR}) = \mu_2$. Clearly we first need to solve \eqref{eq:GH-LR} before proceeding further.

Writing \eqref{eq:GH-LR} as
\begin{equation}
    \bra{L} \dot{\hat{\rho}} \ket{R} = b_{LR} \bra{L} \hat{\rho} \ket{R} + b_{RL} \overline{\bra{L} \hat{\rho} \ket{R}}
\end{equation}
we get a first order differential equation for a complex function that also involves the complex conjugate of said function. The standard approach to solving such equations is to write the complex function as sum of real an imaginary parts and get a system of two first order differential equations, which are solvable using the standard methods. Let us thus denote $f_1 = \Re(\bra{L} \hat{\rho} \ket{R})$ and $f_2 = \Im(\bra{L} \hat{\rho} \ket{R})$, we will use $\vec{f}$ to denote the respective vector containing $f_1$ and $f_2$. Then \eqref{eq:GH-LR} is equivalent to the equation
\begin{equation}
    \dot{\vec{f}} = A \vec{f}
\end{equation}
where
\begin{equation}
    A = 
    \begin{pmatrix}
        \Re(b_{LR} + b_{RL}) & \Im(b_{LR} - b_{RL}) \\
        \Im(b_{LR} + b_{RL}) & \Re(b_{LR} - b_{RL})
    \end{pmatrix}.
\end{equation}
The solutions will now depend on the eigenvalues of the matrix $A$: if $A$ has two real eigenvalues we get the solution of the form $\vec{f} = c_1 \e^{\lambda_1 t} \vec{v_1} + c_2 \e^{\lambda_2 t} \vec{v_2}$, where $\lambda_1$, $\lambda_2$ are the eigenvalues, $\vec{v}_1$, $\vec{v}_2$ are the respective eigenvectors and $c_1$, $c_2$ are the constants to be set from the initial conditions. Since the density matrix must be positive semidefinite, the off-diagonal elements and thus $\vec{f}$ must be bounded and so one should expect $\lambda_1$ and $\lambda_2$ to be negative. If $A$ has two complex roots, then they are of the form $\lambda \pm i \omega$ and the respective eigenvectors $\vec{v}$ and $\vec{\bar{v}}$ are complex conjugates of each other as well. We get the solution of the form $\vec{f} = c_1 \e^{(\lambda + i \omega) t} \vec{v} + c_2 \e^{(\lambda - i \omega) t} \vec{\bar{v}}$ where again $\lambda$ should be negative in order to keep the density matrix positive semidefinite at all times. The last possibility is that $A$ has only one repeated eigenvalue in which case the solution is of the form $\vec{f} = c_1 \e^{\lambda t} \vec{v} + c_2 \e^{\lambda t} (\vec{w} + t \vec{v})$ and again $\lambda$ should be negative. After a concrete solution for $\bra{L} \hat{\rho} \ket{R}$ is obtained, it can be plugged into \eqref{eq:GH-LL-final}, this will become a simple first order ordinary differential equation that can be solved analytically or numerically.

\end{document}